\documentclass[12pt]{article}
\usepackage{epsf,amsmath,amsfonts,bbold}
\usepackage{appendix}
\def\a{\alpha}
\def\b{\beta}
\def\c{\chi}
\def\d{\delta}

\def\eps{\varepsilon}

\def\g{\gamma}

\def\p{\partial}

\def\z{\zeta}

\def\Tr{\mathrm{Tr}}

\newcommand{\bi}{\bibitem}

\def\f{\frac}

\def\be{\begin{equation}}
\def\ee{\end{equation}}

\def\bea{\begin{eqnarray}}
\def\eea{\end{eqnarray}}

\def\ba{\begin{array}}
\def\ea{\end{array}}

\def\bc{\begin{center}}
\def\ec{\end{center}}

\def\bl{\begin{flushleft}}
\def\el{\end{flushleft}}

\def\br{\begin{flushright}}
\def\er{\end{flushright}}

\def\bi{\begin{itemize}}
\def\ei{\end{itemize}}

\def\l{\left}
\def\r{\right}

\def\mc{\mathcal}

\def\bt{\begin{tabular}}
\def\et{\end{tabular}}

\def\nn{\nonumber}

\begin{document}
\begin{flushright}
{FTPI-MINN-11/29} \\
{UMN-TH-3021/11}\\
\end{flushright}

\vspace{2cm}

\renewcommand{\thefootnote}{\arabic{footnote}}
\setcounter{footnote}{0}
\bc
{\Large \bf Cecotti--Fendley--Intriligator--Vafa Index in a Box \\}
\vspace{1cm}

{\large \bf A. Monin$^\dagger$ and M.A.Shifman$^\ddag$} \\

\vspace{0.5cm}

$^\dagger$ {\it \'Ecole Polytechnique F\'ed\'erale de Lausanne, \\
CH-1015 Lausanne, Switzerland}

$^\ddag$ {\it School of Physics and Astronomy, University of Minnesota, \\ Minneapolis, MN
55455, USA} \\
\ec

\vspace{1cm}

\begin{abstract}
The Cecotti--Fendley--Intriligator--Vafa (CFIV) index in two-dimensional ${\mathcal N} =(2,2)$ models
is revisited. We address the problem of ``elementary" (nontopological) excitations over a
kink solution, in the one-kink sector (using the
Wess--Zumino or Landau--Ginzburg models with two vacua as  examples). In other words, we limit ourselves to the large-$\beta$ limit.
The excitation spectrum over the BPS-saturated at the classical level kink is discretized 
through a large box with judiciously chosen boundary conditions. The boundary 
conditions are designed in such a way that half of supersymmetry is preserved as well as
the BPS kink itself, and relevant zero modes. The excitation spectrum acquires a mass gap. All (discretized) excited states enter in four-dimensional multiplets (two bosonic states + two fermionic). Their contribution to $\mathrm{ind} _ {\rm CFIV}$ vanishes level by level. 
The ground state contribution produces $|\mathrm{ind} _ {\rm CFIV}|=1$.
\end{abstract}

\newpage

\section{Introduction}
\label{intro}

In 1982 Witten suggested~\cite{Witten:1982df} in 
supersymmetric theories his famous index 
\be
\Tr (-1)^ F\equiv \sum_n \left\langle n \left | (-1)^ F\right| n\right\rangle,
\ee
where $F$ is the fermion number operator.
Sometimes the Witten index is represented in the form
\be
\mathrm{ind} _{\rm W} = \Tr (-1) ^ F \, e^{ - H \b },
\label{1}
\ee
where the last factor is introduced for ultraviolet (UV) regularization. If the supersymmetric theory 
under consideration is properly regularized 
(i.e. there are no gapless excitations and supersymmetry is unbroken) 
$\mathrm{ind} _{\rm W} $ does not depend on $\beta$ since all nonzero-energy states cancel out
because of the Bose--Fermi degeneracy of the spectrum.
 This can be achieved  on compact spaces, generally speaking.
Then Eq.~(\ref{1}) can be replaced by the original $\Tr (-1) ^ F$.

The crucial feature of  the Witten index is that it is invariant under 
continuous deformations of the initial theory (see Section \ref{gene} below for a more detailed discussion). 
One can deform the theory in any continuous way, in particular, in a way, 
that makes Witten index calculation easy.
The Witten index counts the difference between the number of bosonic and fermionic vacua. Therefore, 
 if $\mathrm{ind} _W \neq 0$  one can conclude that supersymmetry is not spontaneously
 broken. In essence, in four-dimensional theories  with a mass gap $\mathrm{ind} _W $ gives the number 
 of supersymmetric vacua. There exists a vast literature devoted to Witten index in various supersymmetric theories.
 
 A  related ``index" was suggested by Cecotti, Fendley, Intriligator, and Vafa (CFIV)~\cite{Cecotti:1992qh}.
 The CFIV ``index" plays the same role with regard to short soliton (kink) multiplets in ${\mathcal N}=2$
 two-dimensional models as the Witten index for supersymmetric vacua.
 The quotation marks are used to emphasize that the ``index" does not  depend only on a limited class
 of continuos deformations, 
as will be explained shortly.\footnote{ Hereafter the quotation marks will be omitted.} It was introduced 
  as follows:
\be
\mathrm{ind} _ {\rm CFIV} = \Tr \left[ F \, (-1)^F \,e^{ - H \b }\,\right].
\label{2}
\ee
 For any long  supermultiplet
(i.e. the supermultiplet which is not BPS-saturated)
\be
\mathrm{ind} _ {\rm CFIV} \supset f (-1) ^ f + 2 ( f + 1 ) (-1) ^ { f + 1}  + ( f + 2 ) (-1) ^ {f + 2} = 0\,.
\label{3}
\ee
The fermion charges of the states in the
long multiplet are $f+2$, $f+1$, and $f$, respectively, while the corresponding state multiplicities can be read off from 
(\ref{3}). 

However, the  short (BPS-saturated) supermultiplets, if present in the theory, produce a nonvanishing contribution,
\be
\mathrm{ind}_{\rm CFIV} \supset f (-1) ^ f + (f+1) (-1) ^ {f + 1} = (-1) ^ {f+ 1 }\,.
\label{3p}
\ee
Thus,  the value of the CFIV index counts the number of short multiplets that 
cannot be combined into long ones. Again, if $\mathrm{ind}_{\rm CFIV} \neq 0$ is established in an appropriately deformed theory, one can be certain that the original (undeformed theory) supports BPS-saturated solitons. 

The CVIF index does {\em not} depend on 
continuous deformations
of the $D$ terms (i.e. changing the K\"ahler potential). Unlike the Witten index, the
CVIF index can depend, however, on deformations of the $F$ terms \cite{Cecotti:1992qh}
(see Section \ref{gene} for a more detailed discussion). That's why 
in fact it is not  index in the conventional sense,
unlike the Witten index, and that's why we, following the authors, used the quotation marks in the discussion preceding Eq.~(\ref{2}). Nevertheless, the CVIF index is a useful tool, since in some 
important instances
one needs to explore the issue of the BPS saturation under the condition that
the superpotential is exactly known while the K\"ahler potential is not. Then, calculating
the CVIF index, say,  for the canonic K\"ahler potential we may be sure that it stays the same for
any other K\"ahler potential that can be obtained from the canonic one by a continuous deformation.

The regularizing factor $e^{ - H \b }$ is introduced in (\ref{2}) for the same reason as in (\ref{1}). 
In the presence of a continuous spectrum of excitations, isolating and counting distinct supermultiplet contributions 
to the  indices  is a subtle procedure. The notion of degeneracy of the bosonic and fermionic states,
if both belong to the continuous spectra,
 is not well defined because one should include the density of these states in the count.
It is often said that nonzero-energy states in (\ref{1}) and excitations over the BPS kink in (\ref{2}) 
can defy the Bose--Fermi cancellation and produce nonvanishing contributions to the indices due to 
lack of supersymmetry in the {density of the excited states}.

In the pioneering paper~\cite{Cecotti:1992qh} the main emphasis was put on the the multikink sectors. At large $\beta$ they are exponentially suppressed compared to the one-kink sector,
but at finite $\beta$ their contribution is highly nontrivial. In fact, the authors of~\cite{Cecotti:1992qh}
succeeded in obtaining and solving an exact equation for  $\mathrm{ind}_{\rm CFIV} $ as a function of 
$\beta$. A number of models considered in~\cite{Cecotti:1992qh} are integrable, implying, through the thermodynamic Bethe ansatz, an exact $S$ matrix. This knowledge led the authors to the exact
solution.

The problem we address is more limited in scope; it concerns the infrared limit $\beta \to \infty$.
In other words we will focus on the one-kink sector, 
with the goal of obtaining a clear-cut physical interpretation of the role of 
``elementary" (nontopological) excitations over a
kink solution in the process of calculating $\mathrm{ind}_{\rm CFIV} $ in the infrared limit.

Problems arising due to the continuous nature of the 
spectrum of elementary excitations
 is not a unique feature of the CFIV index. The presence of continuum presents a certain difficulty in the calculation of the Witten index too
(see e.g.~\cite{smilga}). A physical way to get rid of this subtlety
is to discretize the would-be continuous spectrum of excitations.
This can be viewed as a technical problem, of course, but its analysis is helpful in understanding related
issues (such as curves of the marginal stability, or `domain walls').

We suggest an infrared regularization (a large box regularization) which discretizes the spectrum of
excitations over the kink under consideration
and maintains enough supersymmetry to provide a 
level-by-level cancellation so that the $\mathrm{ind} _{\rm CFIV}$ is saturated 
by the ground-state supermultiplet.
A very clear-cut physical picture behind the CFIV index calculation emerges.

Before we proceed to the outline of our main idea we hasten to make a reservation.
Since the total spatial momentum is conserved, we can (and will) 
calculate $\mathrm{ind}_{\rm CFIV} $ in
the subspace of the Hilbert space with the vanishing total spatial momentum. The authors 
of~\cite{Cecotti:1992qh} chose not to use this possibility, and integrated over kink's spatial momentum in calculating $\mathrm{ind}_{\rm CFIV} $. Correspondingly, in defining $\mathrm{ind}_{\rm CFIV} $,
they had to use appropriate normalization factors which will be omitted in our analysis. We dot need
them since we do not integrate over kink's spatial momentum. In the limit $\beta\to\infty$ the one-kink state at rest is singled out as the state with the lowest energy for the given boundary conditions.

Our idea is as follows.\footnote{The basic idea presented here
is a generalization of the consideration carried out  previously~\cite{Shifman:1998zy}
in the context of ${\mathcal N} =(1,1)$ theories.}
Assume that the spatial dimension is limited by a large box, with size $L$ and appropriately chosen boundary conditions.
In a sense, the space is compactified, but at the very end we can tend $L\to \infty$. The boundary conditions on the edges of the large box  must be imposed  in such a way that:

(i) they discretize the excitation spectrum;

(ii) they maintain half  supersymmetry; and
 
(iii) they preserve both, the soliton under consideration and fermionic zero modes, which correspond to the broken supercharges.

If this task is achieved, then only the ground state in the given sector will contribute to~(\ref{2}).
Correspondingly, we  obtain (\ref{3}) and (\ref{3p}). 
For all non-BPS (excited) levels the degeneracy is four-fold. Indeed,
two preserved supercharges guarantee doubling of all nonzero energy levels, while the addition of the fermionic zero 
 mode does not change the energy. Denoting by $c _ 0 ^ \dagger$ and  $q ^ \dagger$ the creation operator for the fermionic zero mode and 
 the preserved supercharge, respectively, the long multiplet can be written as
\be
\l | n \r \rangle, q ^ \dagger \l | n \r \rangle, c _ 0 ^ \dagger \l | n \r \rangle, c _ 0 ^ \dagger q ^ \dagger \l | n \r \rangle \,, 
\ee
while the short multiplet is
\be
\l | 0 \r \rangle, c _ 0 ^ \dagger \l | 0 \r \rangle\, .
\ee
One can see that the fermion numbers for such states are indeed as in (\ref{3}). As a result,  
no dependence on $\beta$ will ever appear.

Admittedly, conventional choices of the boundary conditions destroy supersymmetry, and with it is gone 
 degeneracy between 
individual bosonic and fermionic excitation energies. Moreover, ``inappropriate"
 boundary conditions may destroy the kink itself. The straightforward intuitive counting 
 of $\mathrm{ind} _ {\rm CFIV} $, as in (\ref{2}) and (\ref{3}),
becomes impossible,  and one has to invoke the  original CFIV procedure~\cite{Cecotti:1992qh} or similar,
which entangles the continuous spectrum. 

In this paper we will show that the boundary conditions 
satisfying the above conditions (i), (ii) and (iii) exist in two-dimensional ${\mathcal N} =(2,2)$ models -- the subject of the original CFIV analysis --
in much the same way as they had been shown to exist \cite{Shifman:1998zy}  in two-dimensional 
${\mathcal N} =(1,1) $ models. 

Below we will consider as an example the Wess--Zumino models with one or more superfields.
We will  put the system in a large box, preserving the soliton solution, two supercharges, 
and the fermionic zero modes. As a result, the ground states form a short 
supermultiplet, while all (discretized) exited states are in the long supermultiplets.
This allows us to isolate contributions to the CFIV index at each energy level separately.

Why the knowledge of indices is so useful, in particular, the CFIV index?
There exists a number of important problems in which the precise 
form of the K\"ahler potential is not known. For instance, in the CP$(N-1)$ models
the mirror representation exists \cite{mirror}
 which allows one to establish the superpotential, but not the
K\"ahler potential. Assume we want to address the question whether BPS-saturated solitons exist in the
CP$(N-1)$ models. No direct solution for solitons is possible because of strong coupling in the CP$(N-1)$ models.
One can then turn to the mirror representation. Due to the fact that the CFIV index is independent
of the K\"ahler potential, one can use the canonic K\"ahler potential, determine the CFIV index, and 
make the conclusion of the existence of $N$ BPS-saturated solitons. 
This conclusion will stay valid in the CP$(N-1)$ models.

In certain instances a relation between the CFIV index for kinks and Witten index for
an emerging model on the kink world line can be established. The BPS saturation of the kink under consideration is then interpreted as the existence of the supersymmetric vacuum
in quantum mechanics on the world line. The fact that $|\mathrm{ind} _ {\rm CFIV}|$
is integer in our procedure has a clear-cut meaning from this standpoint.

Organization of the paper is as follows. In Section 2 we present the general idea 
as to how to put the system in the box satisfying 
the conditions listed above. In Sections 3 and 4 we describe the general 
${\mathcal N}=2$ system in quadratic approximation and show that the only nonzero contribution comes from the ground state. In Section 5 we consider some examples. In Section 6 we briefly describe generalization to multifield models.

\section{General construction. Putting the system in a box}

To begin with we will consider the system given by the Lagrangian of the form
\bea
\mc{L} & = &  
\f {1} {2} \l ( \p _ R \bar \phi  \p _ L \phi + \p _ L \bar \phi  \p _ R \phi \r ) + \bar F F + 
\mc{W} ' (\phi) F + \bar{\mc{W}} ' (\bar \phi) \bar {F}  \nn \\[2mm]
&+& i \bar \psi _ R  \p _ L \psi _ R + i \bar \psi _ L  \p _ R \psi _ L  - i \mc{W} '' (\phi) \psi _ R \psi _ L  
- i \bar {\mc{W}} '' (\bar\phi) \bar {\psi} _ R \bar {\psi} _L\,,
\label{Lagrangian}
\eea
invariant under the following $\mc{N} = (2,2)$ SUSY transformations:
\bea
\d \phi & = & i \sqrt {2}\l ( \eps _ L \psi _ R + \eps _ R \psi _ L \r ), \nn \\[1mm]
\d \psi _ L & = & - \sqrt {2} \bar \eps _ R \p _ L \phi  - \sqrt {2} \eps _ L F, \nn \\[1mm]
\d \psi _ R & = & - \sqrt {2} \bar \eps _ L \p _ R \phi  + \sqrt {2} \eps _ R F, \nn \\[1mm]
\d F & = & i \sqrt {2} \l ( \bar \eps _ L \p _ R \psi _ L - \bar \eps _ R \p _ L \psi _ R \r ) \,,
\label{transform_c}
\eea
where
\be
\p _ {L,R} = \p _ 0 \pm \p _z.
\ee
If the superpotential $\mc {W}$ has more than one minima there exist solitons interpolating between any two vacua at spatial infinities. Without loss of generality (see Appendix \ref{arb_vac}) we can assume that
\be
\g = \f {\Delta \bar  {\mc {W}}} {\l | \Delta {\mc {W}} \r |} = 1\, ,
\label{9}
\ee
where $\Delta \bar  {\mc {W}} = \bar  {\mc {W}} (\infty) - \bar  {\mc {W}} (-\infty)$. In this case 
the Bogomol'nyi equation for the BPS soliton is given by
\be
\p _ z \phi_k - \bar{ \mc{W}}'(\bar\phi _ k) = 0\, ,
\ee
where the subscript $k$ stands for kink.
A solution to this equation, if it exists,  breaks two supercharges out of four. 
The action of the broken supercharges produces fermionic zero modes,
\bea
\d \psi _ L & = & \sqrt{2} \eta \p _ z \phi _ k, \nn \\[2mm]
\d \psi _ R & = & \sqrt{2} \bar \eta \p _ z \phi _ k \,.
\label{zero_mode}
\eea
There are two real zero modes (or a complex mode and its conjugate).

The supercharge preserving the BPS solution corresponds to a specific linear combination of 
the $\mc{N} = (2,2)$ supercharges, namely,
\be
q = i \sqrt {2} \int d z \l [ \psi _ R \l ( \p _ R \bar \phi + { \mc{W}} ' \r ) 
+ \bar \psi _ L \l ( \p _ L \phi - \bar{ \mc{W}} '  \r ) \r ].
\ee
Needless to say, 
 $q^\dagger$ is conserved too.
In the model under consideration, in addition to supersymmetry, there exists a chiral U(1) symmetry
\bea
\psi _ R ' & = & \psi _ R e ^ {i \a}, \nn \\[2mm]
\psi _ L ' & = & \psi _ L e ^ {- i \a}\,,
\eea
see (\ref{Lagrangian}).
The generator corresponding to this symmetry is the fermion number operator which can be defined as
\be
F = \int d z \l ( \bar \psi _ R \psi _ R - \bar \psi _ L \psi _ L \r ).
\ee
Separating the contribution of the central charge ($Z = \int d z  \, \p _ z \l ( \mc {W} + \bar{\mc {W}} \r )$)  the Hamiltonian $H$
can be written as
\bea
H
&=&
\int dz \,\mc {H} \,,\nonumber\\[2mm]
\mc {H} & = & 
 \p _ z \l ( \mc {W} + \bar{\mc {W}} \r ) 
\nonumber\\[2mm] 
&+&
\p _ 0 \bar \phi \p _ 0 \phi + \l (  \p _ z \phi  - \bar{\mc {W}}'\r ) \l (  \p _ z \bar \phi  - {\mc {W}}'\r )
 \\[2mm]
&+&\f {i} {2} \l ( \bar \psi _ L \p _ z \psi _ L - \bar \psi _ R \p _ z \psi _ R - \p _ z \bar \psi _ L  \psi _ L 
+ \p _ z \bar \psi _ R \psi _ R\r ) + i {\mc {W}}'' \psi _ R \psi _ L  + i \bar{\mc {W}}'' 
\bar \psi _ R \bar \psi _ L \nonumber\,.
\label{hamiltonian}
\eea
Using the expression for relevant operators in terms of fields and  canonic commutation relations
one obtains the following algebra:
\bea
\l [ F, q \r ] & = & - q, \nn \\[2mm]
\l [ F, q ^ \dagger \r ] & = & q ^ \dagger, \nn \\[2mm]
\l [ F, H - Z \r ] & = & - \f {i} {2} \int d z \,\partial_z \l ( \bar \psi _ R \psi _ R + \bar \psi _ L \psi _ L \r ), \nn \\[2mm]
\l [ q, H - Z \r ] & = & - \f {1} {2 \sqrt{2}} \int d z \,\partial_z\l [\bar \psi _ L \l ( \p _ L \phi - \bar{ \mc{W}} '  \r ) - 
\psi _ R \l ( \p _ R \bar \phi + { \mc{W}} '  \r ) \r ].
\label{algebra}
\eea
In order for the fermion operator $F$ and the supercharge $q$ to be conserved in the system in the box, 
one has to require the vanishing of all  boundary terms in (\ref{algebra}). It is evident that the following 
choice of boundary conditions meets this requirement:
\bea
\l ( \bar a, - a \r )
\left.
\l (
\ba{c} 
\psi _ R \\ 
\bar \psi _ L
\ea
\r ) \right| _  { \pm \f {L} { 2 }} & = & 0\,, \nn \\[3mm]
\l ( \bar a, - a \r )
\left.
\l (
\ba{c} 
\p _ L \phi - \bar{ \mc{W}} '  \\ 
\p _ R \bar \phi + { \mc{W}} '
\ea
\r ) \right| _  { \pm \f {L} { 2 }} & = & 0\,,
\label{4bcc}
\eea
where $a$ is an arbitrary complex constant. In Appendix \ref{high_exc} we illustrate how the above boundary conditions work
in the topologically trivial vacuum (i.e. without the kink background).

There are four boundary conditions in (\ref{4bcc}) for the fermions as is required by the first-order differential equations. However, those are the conditions for only two out of four (real) linear combinations of fermions. In order to have boundary conditions for the rest of fermions and at the same time not to overdetermine the system, one has to impose additional boundary conditions which are dependent (trough the equations of motion; see Appendix \ref{extra_bc}). As a result we get
the second set of boundary conditions for the fermion fields,
\bea
\l ( \bar a, - a \r )
\left.
\l ( 
\ba{cc}
\p _ z & - \bar {\mc{W}}''  \\
{\mc{W}}'' & - \p _ z
\ea
\r )
\l (
\ba{c} 
\psi _ R \\ 
\bar \psi _ L
\ea
\r ) \right| _  { \pm \f {L} { 2 }} = 0\, .
\label{2bcc}
\eea

\section{Quadratic approximation and the spectrum}

It should be noted that the boundary conditions introduced above make half supersymmetry manifest 
in all orders of perturbation theory. This is not the end of the story, however. As 
was explained in Section \ref{intro},  we need to preserve the fermionic zero mode as well. 
Here we will demonstrate that  the boundary conditions of the form (\ref{4bcc}) and (\ref{2bcc}) satisfy this 
requirement in the quadratic approximation. The generalization for any order in perturbation theory is given in
Appendix \ref{gen_bc}.

Expanding the Hamiltonian around the BPS background we arrive at
\bea
\l [ \mathcal{H} - \p _ z \l ( \mc {W} + \bar{\mc {W}} \r ) \r ] _ \text{quad} 
\!\!\!
&=&  
\!\!\!\f {1} {2} \l ( \p _ 0 \bar\c, \p _ 0 \c \r ) 
\l (
\ba{c} 
\p _ 0 \c \\ 
\p _ 0 \bar \c
\ea
\r ) \nn \\[3mm]
& + &
\f {1} {2} \l ( \bar\c, \c \r ) P ^ 2 
\l (
\ba{c} 
\c \\ 
\bar \c
\ea
\r )
+
\f {1} {2} \l ( \bar \psi _ L, \psi _ R \r ) P 
\l (
\ba{c} 
\psi _ L \\ 
\bar \psi _ R
\ea
\r )
 \nn \\[3mm]
&-&
\f {1} {2} \l ( \bar \psi _ R, \psi _ L \r ) P 
\l (
\ba{c} 
\psi _ R \\ 
\bar \psi _ L
\ea
\r )
-
\f {i} {2} \p _ z \l [\l ( \bar\c, -\c \r ) P 
\l (
\ba{c} 
\c \\ 
\bar \c
\ea
\r )
\r ],
\label{Hamiltonian}
\eea
where
$ \phi \equiv \phi_k +\chi\,,$ and
\bea
P &=&
\l ( 
\ba{cc}
i \p _z & - i \bar {\mc{W}}''  \\
i {\mc{W}}'' & - i \p _z
\ea
\r ), ~~~ P ^ \dagger  = P \nn \\[3mm]
P ^ 2 &=& \l ( 
\ba{cc}
- \p _z ^ 2 + \bar {\mc{W}}'' {\mc{W}}'' & \p _ z \bar {\mc{W}}''  \\
\p _ z {\mc{W}}'' & - \p _z ^ 2 + \bar {\mc{W}}'' {\mc{W}}''
\ea
\r ).
\label{oper_P}
\eea
The fermionic zero mode (\ref{zero_mode}) wich satisfies
\be
P \l (
\ba{c} 
\psi _ R \\ 
\bar \psi _ L
\ea
\r ) = 0,
\ee
is preserved by the choice 
\be
a = \p _ z \phi _ k
\ee
 which implies the linearized boundary conditions
\bea
\left. 
\l ( \p _ z \bar \phi _ k, - \p _ z \phi _ k \r ) P 
\l (
\ba{c} 
\c \\ 
\bar \c
\ea
\r ) \right| _  { \pm \f {L} { 2 }} & = & 0 \, , \nn \\[1mm]
\left.
\l ( \p _ z \bar \phi _ k, - \p _ z \phi _ k \r )
\l (
\ba{c} 
\c \\ 
\bar \c
\ea
\r ) \right| _  { \pm \f {L} { 2 }} & = & 0 \, , \nn \\[1mm]
\left.
\l ( \p _ z \bar \phi _ k, - \p _ z \phi _ k \r ) P 
\l (
\ba{c} 
\psi _ R \\ 
\bar \psi _ L
\ea
\r ) \right| _  { \pm \f {L} { 2 }} & = & 0 \,, \nn \\[1mm]
\left.
\l ( \p _ z \bar \phi _ k, - \p _ z \phi _ k \r )
\l (
\ba{c} 
\psi _ R \\ 
\bar \psi _ L
\ea
\r ) \right| _  { \pm \f {L} { 2 }} & = & 0 \, .
\label{linear_bc}
\eea
The geometrical meaning of the relations above is the following. If we consider the 
$\mathbb{C} ^ 2$ space with coordinates $( z_ 1, z _ 2 )$, then any $\phi _ k$ 
solution defines a subspace (line) in it $$\g _ k = ( \phi _ k, \bar \phi _ k )\,.$$ 
As a result, the boundary conditions (\ref{linear_bc}) are the orthogonality conditions 
(at the boundary $\pm L / 2 $) of the vector normal to $\g _ k$ and the fluctuations.

Now, we expand the fields in a series of eigenfunctions of the operator $P^2$
\bea
\l (
\ba{c} 
\c \\ 
\bar \c
\ea
\r ) & = & \sum _ {n,s} b _ {n s} 
\l (
\ba{c} 
f _ n ^ s \\ 
\bar f _ n ^ s
\ea
\r ) \nn \\[3mm]
\l (
\ba{c} 
\psi _ L \\ 
\bar \psi _ R
\ea
\r ) & = & \sum _ {n,s} \xi _ {n s} 
\l (
\ba{c} 
f _ n ^ s \\ 
\bar f _ n ^ s
\ea
\r ),
\label{expansion}
\eea
where the index $n$ labels the level corresponding to the eigenvalue $\omega _ n$ while $s=1,2$ labels the eigenstates of the operator $P$ with positive and negative eigenvalues correspondingly. The functions $f _ n ^ s$ are such that
\bea
\sum _ {n, s} f _ n ^ s (x) \bar f _ n ^ s (y) & = & \d (x-y) \, , \nn \\[1mm]
\int d z \bar f _ n ^ s (z) f _ m ^ r (z) & = & \f {1} {2} \d _ {n m} \d _ {s r} \,
\label{f_prop}
\eea
and they satisfy the boundary conditions
\be
\l.
\l ( \p _ z \bar \phi _ k, - \p _ z \phi _ k \r ) 
\l (
\ba{c} 
f _ n ^ s \\ 
\bar f _ n ^ s
\ea
\r ) 
\r | _  { \pm \f {L} { 2 }} =
\l.
\l ( \p _ z \bar \phi _ k, - \p _ z \phi _ k \r ) P 
\l (
\ba{c} 
f _ n ^ s \\ 
\bar f _ n ^ s
\ea
\r ) 
\r | _  { \pm \f {L} { 2 }} = 0.
\label{f_bc}
\ee
Plugging the expansion (\ref{expansion}) to the Hamiltonian (\ref{Hamiltonian}) one gets
\bea
H - Z = \f {1} {2} \dot b _ 0 ^ 2 + \f {1} {2} \sum _ {n \neq 0, s} 
\left(
\dot b _ {n s} ^ 2 + \omega _ n ^ 2 b _ {n s} ^ 2 
\right)
+
\sum _ {n \neq 0} \omega _ n \l ( \bar \xi _ {n 1} \xi _ {n 1} + \xi _ {n2} \bar  \xi _ {n2} \r ),
\eea
which upon the change of the variables
\bea
a _ {ns} & = & \f {\omega _ n ^ {1/2} b _ {n s} - i \omega _ n ^ { - 1/2} \dot b _ {n s} } {\sqrt{2}}, \nn \\
\xi _ {n 1} & = & c _ {n1}, \nn \\
\xi _ {n 2} & = & c ^ \dagger _ {n2}, \nn \\
\xi _ 0 & = & c _ 0 ^ \dagger,
\eea
gives
\be
H - Z = \f {1} {2} \dot b _ 0 ^ 2 + \sum _ {n \neq 0, s} \omega _ n 
\l ( a ^ \dagger _ {n s} a _ {n s} + c ^ \dagger _ {n s} c _ {n s} \r ).
\label{Hamiltonian _quad}
\ee

\section{The CFIV index}

\subsection{Fermion charge}

We start from the following remark. 
As one can see from the expression for the Hamiltonian (\ref{Hamiltonian _quad}),
 there is an additional doubling of energy levels (four operators $a _ {ns}$ and $c _ {ns}$). 
 
The expression for the fermion number operator  in terms of creation-annihilation operators
takes the form
\be
F = \f {1} {2} \l ( c _ 0 ^ \dagger c _ 0 - c _ 0 c _ 0 ^ \dagger \r ) 
+ \sum _ {n \neq 0} \l ( c _ {n 2} ^ \dagger c _ { n 2 } - c _ {n 1} ^ \dagger c _ { n 1 } \r ).
\ee
Therefore the fermions of type $c _ {n1}$ have charge $-1$ while those of type  $c _ {n2}$ have charge $1$. 
The fermions produced by $c _ 0 ^ \dagger$ have half-integer fermion charge due to the charge 
fractionalization~\cite{Jackiw:1975fn}.

\subsection{Index}

Choosing an arbitrary excited (non-BPS) state
\be
\l | n \r \rangle = a _ {js} ^ \dagger c _ {jr} ^ \dagger \dots \l | 0 \r \rangle,
\ee
one finds the long multiplet in the form\footnote{We assume the $q ^ \dagger$ does not annihilate the state $\l | n \r \rangle$.}
\be
\l | n \r \rangle, q ^ \dagger \l | n \r \rangle, c _ 0 ^ \dagger \l | n \r \rangle, c _ 0 ^ \dagger q ^ \dagger \l | n \r \rangle, 
\label{32}
\ee
whose contribution to the CFIV index vanishes. The only nonzero contribution to the index is from the short multiplet,
\be
\l | 0 \r \rangle, c _ 0 ^ \dagger \l | 0 \r \rangle,
\ee
which gives
\be
\mathrm{ind} _ {\rm CFIV} = \l ( \f {1} {2} \r ) (-1) ^ {1/2} + \l ( - \f {1} {2} \r ) (-1) ^ {-1/2} = i
\ee

Note that
 any additional  (i.e. not required by the preserved two supercharges) level doubling (at the quadratic level), as
 indicated in the previous subsection,  makes the level in question
effectively $\mc {N} = (2,2)$ supersymmetric. Four-dimensional multiplet (\ref{32}) 
is accompanied by another four-dimensional multiplet with the same energy.
 The contribution to the index from such multiplets vanishes automatically. However, it is not clear whether 
 or not this  latter doubling persists in higher orders. 
 
 Moreover, even if it persists, the ``other" long multiplets (other than (\ref{32})), taken individually,
  {\it do} contribute to the index  see Eq. (\ref{C_q_multiplet}). This is because they are not genuine supermultiplets:
  the fermion charges defy Eq. (\ref{3}).

\section{Examples}

\subsection{Superpolynomial model}

For the polynomial superpotential with real coefficients
\be
\mc {W} \l ( \Phi\r ) = \f {m^ 2} {4 \lambda} \Phi - \f {\lambda} {3} \Phi ^ 3,
\label{super_potential_real}
\ee
the BPS kink solution is given by the following expression
\be
\phi _ k (z) = \f {m} {2 \lambda} \tanh \f {m} {2} z\,.
\label{rsolit}
\ee
It is purely real. The Hamiltonian in quadratic approximation can be written as
follows:\footnote{The following change of the variables was performed
\bea
\c & = & \c _ 1 + i \c _ 2, \nn \\[2mm]
\psi _ R & = & \f {\psi _ 2 - \psi _ 1} {\sqrt{2}}, ~~~ \psi _ 1 = u _ 1 + i u _ 2, \nn \\[3mm]
\psi _ L & = & \f {\psi _ 2 + \psi _ 1} {\sqrt{2}}, ~~~\psi _ 2 = v _ 1 + i v _ 2\nn\,.
\eea}
\bea
\l [ \mc {H} - \p _ z \l ( \mc {W} + \bar{\mc {W}} \r ) \r ] _ {\text {quad} }& = & \p _ 0  \c _ 1 \p _ 0 \c _1 + \p _ 0  \c _ 2 \p _ 0 \c _2  \\ 
& + & \c _ 1 p ^ \dagger p \c _ 1 + i v _ 2 p u _ 2 + i u _ 1 p v _ 1 + \c _ 2 p p ^ \dagger \c _ 2 - i v_ 1 p ^ \dagger u _ 1 - i u _ 2 p ^ \dagger v _ 2, \nn
\label{real_quad}
\eea
where the operator $p$ is defined by
\be
p = \p _ z - \mc{W}'' (\phi _ k) =\p _ z + 2 \lambda \phi _ k\,,
\ee
and the boundary conditions (\ref{linear_bc}) take the form
\bea
p \c _ 1 \Big |_{z = \pm \f {L} {2}} = u _ 1 \Big |_{z = \pm \f {L} {2}} =
v _ 2 \Big |_{z = \pm \f {L} {2}} = \c _ 2 \Big |_{z = \pm \f {L} {2}} = 
p u _ 2 \Big |_{z = \pm \f {L} {2}} = p v _ 1 \Big |_{z = \pm \f {L} {2}} = 0\,.
\eea
The operators $p ^ \dagger p$ and  $p  p ^ \dagger$ have the same eigenvalues 
(except zero) and their 
eigenfunctions are related by\,\footnote{The modes satisfy the  
boundary conditions $\tilde f \Big |_{z = \pm \f {L} {2}} = 0$ and  $\ p f \Big |_{z = \pm \f {L} {2}} = 0\,.$}
\bea
\tilde f _ n &=& \f {1} {\omega _ n} p f _ n, \nn \\[3mm]
f _ n &=& \f {1} {\omega _ n} p ^ \dagger \tilde f _ n,
\eea
except for the zero mode of the operator $p ^ \dagger p$,
\be
p f _ 0 = 0.
\ee
The expansion in series in eigenfunctions leads to the same result as described above. 

\subsection{CP(1) mirror}

Another example is the system with the superpotential appearing as a mirror in
the CP(1) model \cite{mirror}
\be
\mc {W} = \f {\lambda} {2} \l ( \Phi + \f {v ^ 2} {\Phi} \r ).
\ee
There are two BPS kinks in this case corresponding to two semicircles,  in the upper and lower complex half planes,
\be
\phi _ k ^ {1,2} = v \l ( \tanh \f {\lambda z} {v} \pm i \, {\cosh ^ {-1} \f {\lambda z} {v} } \r ).
\ee
Then we repeat consideration of the previous subsection.

\subsection{C conjugation}

For  systems such as that described by the superpotential (\ref{super_potential_real}),
 with all real coefficients, there is a charge conjugation symmetry $C$
\bea
C \phi C & = & \bar \phi\,, \nn \\[1mm]
C \psi C & = & \bar \psi\,,
\eea
which is not spontaneously broken by the kink solution (\ref{rsolit}). Therefore, the following 
excited non-BPS states are degenerate:
\be
\l | n \r \rangle, q ^ \dagger \l | n \r \rangle, C q ^ \dagger \l | n \r \rangle, C \l | n \r \rangle.
\label{C_q_multiplet}
\ee
The fermion charge assignments are, naturally, different from those we used in (\ref{3}).
Indeed, due to the fact that
\be
C F C = - F,
\ee
the contribution to the index of the multiplet (\ref{C_q_multiplet})
\be
f  (-1) ^ {f} + \l ( f + 1 \r )(-1) ^ {f + 1} - f  (-1) ^ {- f} - \l ( f + 1 \r )(-1) ^ {- f - 1} = (-1) ^ {-f} - (-1) ^ {f}
\ee
does not vanish unless $f=1$. Only if one adds the multiplet (\ref{C_q_multiplet}) to the degenerate
one, with the fermion state on the zero-energy level\footnote{For which the fermion number is $f+1$.},
 does one get the overall zero contribution.

\section{Multifield Wess--Zumino models}
\label{mwzm}

Our consideration can be easily generalized for the case of more than one field.
 Let us briefly sketch the procedure focusing on the bosonic fields. 
 For $n$ fields there is a contribution to the Hamiltonian of the form
\be
\mathcal{H} - \z ^ {00} \supset \l (  \p _ z \phi _ i  - \bar{\mc {W}} _ i \r ) \l (  \p _ z \bar \phi _ i  - {\mc {W}} _ i \r ),
\ee
where $$\mc {W} _ i = \f {\p \mc {W} } {\p \phi _ i}\,,\quad i= 1,2, ..., n\,.$$ Therefore, 
the kink solution satisfies $n$ equations
\be
\p _ z \phi _ i ^ k - \bar{\mc {W}} _ i (\bar \phi ^ k) = 0\,.
\ee
Upon expansion around the kink the Hamiltonian becomes
\be
\l ( \mathcal{H} - \z ^ {00} \r ) _ {\text{quad}} \supset 
\f {1} {2} \l ( \bar\c, \c \r ) P ^ 2 
\l (
\ba{c} 
\c \\ 
\bar \c
\ea
\r ),
\ee
where now $\c$ and $\bar \c$ are the columns of $n$ elements
\bea
\c = \l (
\ba{c} 
\c _ 1 \\ 
\vdots \\
\c _n
\ea
\r ), ~~~
\bar \c = \l (
\ba{c} 
\bar \c _ 1 \\ 
\vdots \\
\bar \c _n
\ea
\r ),
\eea
and the operator $P$ has the form
\bea
P = \l ( 
\ba{cc}
i \delta _ {i j} \p _z & - i \bar {\mc{W}} _ {i j}   \\
i {\mc{W}} _ {i j}  & - i \delta _ {i j}  \p _z
\ea
\r ),
\eea
with self-evident notation.

In order to discretize the spectrum we have to impose a boundary conditions for each field.\footnote{For $n$ complex fields we have to impose $2n$ boundary conditions. For one field we had two.} We will act in the same way as before. We introduce the $2n$-dimensional complex space $\mathbb{C} ^ {2 n}$ with a scalar product
\be
(w,z) = w ^ \dagger z.
\ee
For the $n$-dimensional subspace consisting of all points $(z, \bar z)$, there is an induced scalar product
\be
(w,z) = w ^ \dagger z + z ^ \dagger w\, ,
\ee
which is real and has a usual form if one introduces real coordinates $$z _ i = \f{ x _ i + i y _ i } { \sqrt {2} }\,,$$ namely,
\be
(w,z) = u _ i x _ i + v _ i y _ i\, .
\ee
The kink can be represented as a line in this $2n$-dimensional hypersurface,  
\be
\gamma _ n ^ k = (\p _ z \phi _ 1 ^ k, \dots, \p _ z \phi _ n ^ k, \p _ z \bar \phi _ 1 ^ k, \dots, \p _ z \bar \phi _ n ^ k)\,.
\ee
Then the boundary conditions are in fact the orthogonality conditions 
between the norms to the curve $\g _ n ^ k$ and the fluctuations.

\section{A general perspective}
\label{gene}

Now, after we finished our box construction, we would like to discuss issues common to
the Witten and CFIV indices from a more general standpoint.

The statement that the Witten index does not depend on
the continuous deformations of the superpotential is
a  mathematically rigorous assertion. However, from the physics standpoint 
this assertion should be qualified. Indeed, it may well happen that
under continuous deformations of the superpotential a supersymmetric vacuum (or vacua)
of the theory run away to infinity in the space of fields, while
a nonsupersymmetric minimum remains near the origin. This means that the supersymmetric vacuum
decouples from the physical Hilbert space implying a change in the Witten index evaluated in
the physical Hilbert space. The most well-known example of this type
is the
 Intriligator--Thomas--Izawa--Yanagida (ITIY) model \cite{IT}:
 an SU(2) super-Yang--Mills theory with a judiciously chosen matter sector
 (for a review see \cite{AV}).
 
The model is nonchiral, therefore, the Witten  index equals two. Nevertheless, 
supersymmetry is dynamically broken, i.e. effectively the Witten index vanishes!

In the ITIY model we have four ``quark" superfields $Q^{\alpha}_ f$, each is 
a color doublet ($\alpha = 1,2$ and $f= 1,2,3,4$).   
  In addition to the  quark  superfields 
$Q^{\alpha}_ f$,  six color-singlet chiral
superfields $S^{fg}=-S^{gf}$ are introduced.  Their 
 interaction  with  $Q_{\alpha f}$ is due to the superpotential, 
\begin{equation}
{\cal W} = \frac{h}{2} \, S^{fg} \,Q^{\alpha}_ f \, 
Q^{\beta}_g\,\epsilon_{\alpha
\beta} +m S^2\;.
\label{spit}
\end{equation}
Two supersymmetric vacua were found \cite{IT} at
\begin{equation}
S = \pm \mbox{const}\, h \, m^{-1} \Lambda^2\, , 
\end{equation}
in full accord with Witten's index. However,
in the limit   $m\to 0$, when the second term in (\ref{spit}) 
disappears  (certainly a smooth allowed deformation of the superpotential)
 these supersymmetric vacua escape to infinity in the space of fields. 
A non-supersymmetric vacuum survives  at a finite distance from the origin in the space of fields \cite{IT}.
From the physical point of view, in passing from
$m\neq 0$ to $m=0$ the Witten index jumps by two units.

The statement that the CFIV index is independent of the
continuous deformations of the  K\"ahler potential but depends on
deformations of the superpotential, being mathematically 
accurate, must be qualified too. Indeed, if we have a short kink supermultiplet, and 
the superpotential parameters are not in the immediate vicinity of 
the curves (walls) of marginal stability, small variations of the superpotential
cannot make a long supermultiplet out of the short one. 
Only if we change the parameters in such a way that we touch the
curves (walls) of marginal stability, the missing states (needed to make a long supermultiplet
from the short supermultiplet)
come from the spatial infinity (now we mean not the space of fields, but just
the $z$ axis), see e.g. \cite{ritz}.

When we introduce a large box, strictly speaking, before taking the limit 
$L\to \infty$, we do not have spatial infinities.
If we include the edges of the box into consideration
and will not discriminate between the states
localized on the kink and those localized on the box edges,
all supermultiplets will become long, and the CFIV index of this expanded system 
will vanish regardless of which side of the curve (wall) of marginal stability we are.

Therefore, in both cases discussed above there is a subtlety associated
with the run-away situations: either in the space of fields or in the configurational space.
It is desirable to make the formal index analysis ``know" about possible run-aways,
in the most general form. In our problem we managed to avoid this issue by imposing 
special boundary conditions, a construction which is is obviously not general.

\section{Conclusions}

Our basic idea is straightforward. Discretizing excitation spectrum while preserving enough  
supersymmetry, along with the BPS soliton with its moduli, allows us to achieve the 
nonzero mode cancelation in the CFIV index
level-by-level. With the choice of the boundary conditions as given above in our paper,
supersymmetry is manifest. The mode degeneracy appears in much the same way as in the problem 
of instantons
\cite{instanton}. As a result the calculation of the CFIV index reduces to finding the contribution 
only from the ground state. Therefore, the index can be used to count the number of short 
multiplets in the theory. 


In general,  introducing boundaries or certain conditions far from the kink core 
may change the solution. But physically these possible changes do not affect the kink core per se, 
but may add (or subtract) ``junk" at the boundaries which has to be eliminated from
any physically sensible kink analysis anyway. The box we suggest is subtle, no ``junk" sticks to its edges.

Assume we impose {\it ad hoc} boundary conditions in the kink sector
which need {\it not} maintain supersymmetry. We will require, however, that they preserve the 
BPS kink and discretize the
excitation spectrum, creating a mass gap $1/L$. Then the calculation of the CFIV index
(defined in the normalization we use)
should
produce an integer result coinciding with ours in the limit when we first fix $L$, then 
tend $\beta\to \infty$ and only at the very end allow $L$ to become infinite.
And it does (see Sect. 4
in \cite{Cecotti:1992qh} which can be adjusted to yield this result).
If one is not careful, one can first take $L\to \infty$, which makes the spectrum
continuous and adds to the index some nonvanishing infrared excitation
contributions with energies lower than $1/\beta$. However,
 the limits $L\to \infty$ and $\beta \to \infty$
are not interchangeable.

The procedure we suggest seems natural in view of the fact that
in some instances the calculation of a nonvanishing CFIV index for a given kink is
essentially the same as the calculation of a nonvanishing Witten index in supersymmetric 
quantum mechanics on the kink world line. One of simple examples of this type is
provided by kinks in the ${\mathcal N} = (2,2)$ CP(1) model with a (large) twisted mass. 
It is discussed in detail e.g. in \cite{SY}.

\section*{Acknowledgments}

We are grateful to S. Cecotti, K. Intriligator, C. Vafa, M. Voloshin and A. Vainshtein for useful discussions.
The work of M.S. is supported in part by DOE grant DE-FG02- 94ER-40823 at the University of Minnesota. The work of A.M. is supported 
 in part by Swiss National Science Foundation, FASI RF 14.740.11.0347 (2009-2013) and RFBR 10-02-00509.

\newpage

\appendix

\appendixpage

\section{More on the boundary conditions}

\subsection{Trivial vacuum \label{high_exc}}
In this section we illustrate how the boundary conditions (\ref{4bcc}) and (\ref{2bcc}) or their linear analogs (\ref{linear_bc}) work in the case of topologically trivial vacuum. At the same time it is evident that similar situation occurs even for nontrivial background for sufficiently high excitations. It is enough to consider the eigenfunctions of the operator $P^2$ (\ref{oper_P}) satisfying the boundary conditions (\ref{f_bc}). For the case at hand the operator $P^2$ has the following form
\bea
P ^ 2 = - \p _ z ^ 2 \mathbb{1}.
\eea
To further simplify things we take $a$ from  (\ref{4bcc}) and (\ref{2bcc}) to be real (for high excitations around kink configuration it is always possible to do by simple phase rotation). The general solution for the equation
\be
 - \p _ z ^ 2
\l (
\ba{c} 
f  \\ 
\bar f 
\ea
\r )  = \omega ^ 2 
\l (
\ba{c} 
f  \\ 
\bar f 
\ea
\r ),
\ee
with boundary conditions
\bea
\l .\mathrm{Im} f \r | _  { \pm \f {L} { 2 }} & = & 0, \nn \\
\l . \p _ z \mathrm{Re} f \r | _  { \pm \f {L} { 2 }} & = & 0.
\eea
is given by
\be
f _ n = A \cos \omega _ n \l ( z + \f {L} {2} \r ) + i B \sin \omega _ n \l ( z + \f {L} {2} \r ),
\ee
with $A$ and $B$ being real, while $\omega _ n = \f {\pi n} {L}$. As a result one finds that the functions having the properties
(\ref{f_prop}) are given by the following expressions
\bea
f _ n ^ {1} & = & \f {1} {\sqrt{2 L}} \, e ^ {- i \omega _ n \l ( z + \f {L} {2} \r )}, \nn \\
f _ n ^ {2} & = & \f {1} {\sqrt{2 L}} \, e ^ {i \omega _ n \l ( z + \f {L} {2} \r )}.
\eea

\subsection{Beyond quadratic approximation 
\label{gen_bc}}

We have shown before that the boundary conditions (\ref{4bcc}) and (\ref{2bcc}) are exactly the ones described in Introduction, namely, such boundary conditions discretize the spectrum while preserving the BPS soliton with its modules and half of the supersymmetry. The valid question is, whether such boundary conditions can be generalized for arbitrary order in perturbation theory. To answer this we note that the boundary conditions (\ref{4bcc}) and (\ref{2bcc}) preserve half of the supersymmetry and BPS solution in arbitrary order, while if taken with $a = \p _ z \phi _ k$ they do not respect the fermionic zero mode beyond quadratic approximation. Indeed, the zero mode satisfies the following operator equation
\be
\left\langle
\l ( 
\ba{cc}
\p _ z & - \bar {\mc{W}}''  \\
{\mc{W}}'' & - \p _ z
\ea
\r )
\l (
\ba{c} 
\psi _ R \\ 
\bar \psi _ L
\ea
\r )
\right\rangle = 0,
\ee
where the average is taken with respect to vacuum state (soliton) plus the fermionic mode
\be
\langle \dots \rangle \equiv \langle \text{sol, f} | \dots | \text{sol, f} \rangle.
\ee
Therefore, the second set of boundary conditions (\ref{2bcc}) is automatically satisfied for the fermionic zero mode.
However, we know that the the zero mode is related to the BPS soliton profile through the supersymmetry transformations\footnote{The leading order version of this relations is (\ref{zero_mode}).}
\bea
\langle \d \psi _ R \rangle & = & \bar \eta \p _ z \langle \phi \rangle, \nn \\
\langle \d \psi _ L \rangle & = & \eta \p _ z \langle \phi \rangle.
\eea
Hence, it does not satisfies the (\ref{4bcc}) with $a = \p _ z \phi _ k$, but rather with $a = \p _ z \langle \phi \rangle $. The profile $ \langle \phi \rangle $ can be found order by order in perturbation theory. Therefore, we have found the necessary boundary conditions for arbitrary order in perturbation theory.

\subsection{Dependent boundary conditions \label{extra_bc}}
Consider the system of two first order linear differential equations on the interval $z \in \l [ 0, L\r ]$
\bea
f _ 1 ' & = & -\omega f _ 2, \nn \\
f _ 2 ' & = & \omega f _ 1.
\label{ind_bc}
\eea
In order to make the spectrum discreet one has to impose two independent boundary conditions. The following boundary conditions are possible, since they are independent
\bea
f _ 2 (0) & = & f _ 2 (L) = 0, \nn \\
& \text{or} & \nn \\
f _ 2 (0) & = & f ' _ 1 (L) = 0.
\eea
While the boundary conditions of the form
\bea
f _ 2 (0) & = & f ' _ 1 (0) = 0
\eea
are not independent and therefore can not be used to discretize the spectrum. Suppose we choose the boundary conditions given in the first line of (\ref{ind_bc}). It is clear that adding another boundary conditions like
\bea
f _ 1 ' (0) & = & f ' _ 1 (L) = 0
\eea
does not change the spectrum, since those boundary conditions follow from the first ones and the equations.

\section{Interpolating between arbitrary vacua \label{arb_vac}}

\renewcommand{\theequation}{B.\arabic{equation}}
\setcounter{equation}{0}

For a general parameter
\be
\g = \f {\Delta \bar  {\mc {W}}} {\l | \Delta {\mc {W}} \r |} = e ^ {- 2 i \a},
\ee
after redefining the fermionic fields
\be
\psi _ {R, L} \to e ^ {- i \a} \psi _ {R, L}\,.
\ee
The Hamiltonian can be put in the form  
\bea
\mc {H} & = & \p _ 0 \bar \phi \p _ 0 \phi + \l (  \p _ z \phi  - \bar \g \bar{\mc {W}}'\r ) \l (  \p _ z \bar \phi  - \g {\mc {W}}'\r )
+ \p _ z \l ( \g \mc {W} + \bar \g \bar{\mc {W}} \r )  \\[2mm]
&+&\f {i} {2} \l ( \bar \psi _ L \p _ z \psi _ L - \bar \psi _ R \p _ z \psi _ R - \p _ z \bar \psi _ L  \psi _ L 
+ \p _ z \bar \psi _ R \psi _ R\r ) + i \g {\mc {W}}'' \psi _ R \psi _ L  + i \bar \g \bar{\mc {W}}'' 
\bar \psi _ R \bar \psi _ L \nonumber\,, 
\eea
while the supertransformations upon the  substitution
\be
\eps _ {R, L} \to e ^ {i \a} \eps _ {R, L}
\ee
become
\bea
\d \phi & = & i \sqrt {2}\l ( \eps _ L \psi _ R + \eps _ R \psi _ L \r ), \nn \\[2mm]
\d \psi _ L & = & - \sqrt {2} \bar \eps _ R \p _ L \phi  + \sqrt {2} \eps _ L \bar \g \bar{\mc {W}}', \nn \\[2mm]
\d \psi _ R & = & - \sqrt {2} \bar \eps _ L \p _ R \phi  - \sqrt {2} \eps _ R \bar \g \bar{\mc {W}}'.
\label{transform_c_a}
\eea
Therefore, the problem is reduced to the previously solved problem with superpotential $\g {\mc {W}}$.

\end{document}